\documentclass[aip,reprint]{revtex4-1}
\usepackage{graphicx}
\usepackage{amsmath}
\usepackage{ulem}
\usepackage{hyperref}

\begin{document}
\author{Ajay Muralidharan}
\email{amuralidhar4@wisc.edu}
\affiliation{Department of Chemistry, University of Wisconsin-Madison, Madison, WI 53706}
\author{Arun Yethiraj}
\email{yethiraj@wisc.edu}
\affiliation{Department of Chemistry, University of Wisconsin-Madison, Madison, WI 53706}
\title{Fast estimation of ion-pairing for screening electrolytes: A cluster can approximate a bulk liquid}

\begin{abstract}
The propensity for ion-pairing can often dictate the thermodynamic and kinetic properties of electrolyte solutions. Fast and accurate estimates of ion-pairing can thus be extremely valuable for supplementing design and screening efforts for novel electrolytes. Here, we introduce an efficient cluster model to estimate the local ion-pair potential-of-mean-force (PMF) between ionic solutes in electrolytes. The model incorporates an ion-pair and a few layers of explicit solvent in a gas-phase cluster and leverages an enhanced sampling approach to achieve high efficiency and accuracy. We employ harmonic restraints to prevent solvent escape from the cluster and restrict sampling of large inter ion distances. We develop a Cluster Ion-Pair Sampling (CLIPS) tool that implements our cluster model and demonstrate its potential utility for screening simple and  poly-electrolyte systems. 

\end{abstract}

\pagebreak

\maketitle 
\section{Introduction}

Ion-ion correlations and pair potentials-of-mean-force (PMF) are fundamental elements that enter the theory for ion-pairing in electrolyte solutions.\cite{pratt1994ion,schenter2003generalized,roy2017marcus} The propensity for ion-pairing can often dictate the conductivity and viscosity of electrolyte solutions.\cite{webber_conductivity_1991} In battery electrolytes, for example, weak correlations between lithium ion (Li$^+$) and the anion is often a prerequisite for achieving both a high Li$^+$ conductivity and a high transference number.\cite{fong2021ion} This is because long-lived charge-neutral aggregates do not contribute to useful transport in batteries. In addition, high energetic barriers for dissociation of ions can also lead to lowered salt solubility leading to poor performance. Hence, quick and accurate estimates of ion-pairing and dissociation behavior between ionic species can be an immensely useful criteria for assessing electrolytes. 

\begin{figure}[ht!]
	\includegraphics[width=1\linewidth]{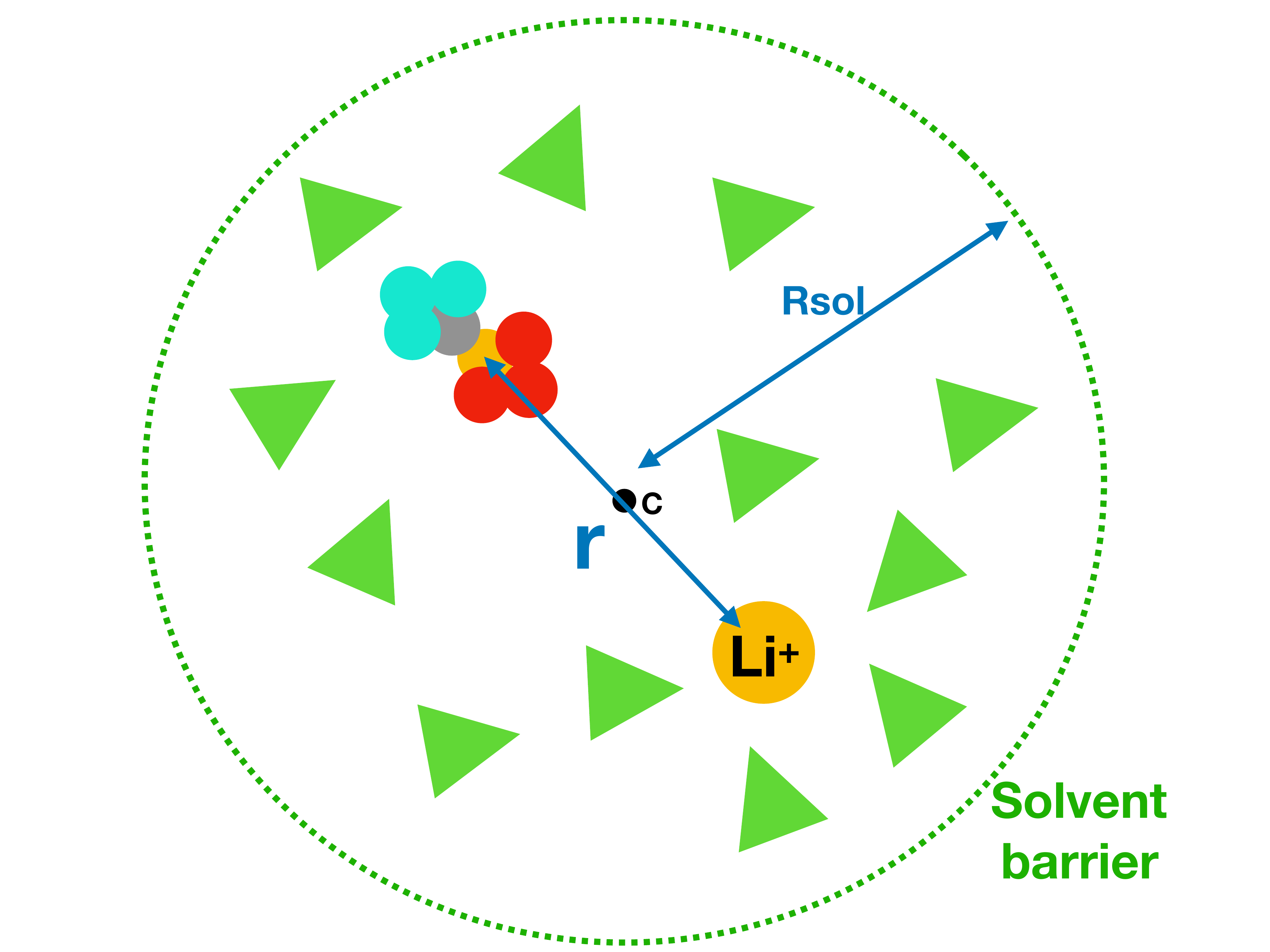}%
	\caption{Schematic representation of the cluster model for estimating the dissociation barrier and short-range PMF between ions. Li$^+$ with arbitrary anion and solvent (green triangles) is pictured above. We perform enhanced sampling along the distance coordinate, $r$, between the ions. A one sided harmonic upper wall restricts sampling of $r$ above 0.65 nm and a solvent barrier prevents solvent diffusion away from the ions.}
	\label{figure:TOOL}%
\end{figure}

 Extensive theoretical and computational efforts have been directed towards estimating ion-ion PMF's in a variety of contexts.\cite{lytle2021lithium,van2016water,zhang2017potential,singh2020driving,fennell2009ion,luo2013simulation,lyubartsev1995calculation,allen2006molecular,kovalenko1999potential} However, obtaining accurate and converged PMF's with computational models that include bulk explicit solvent can be computationally intensive. Because of the difficulty of these molecular calculations, a cluster model for estimating ion pair PMF's that reliably captures the important local features would be extremely valuable. Here, we utilize a simplified cluster model and perform free energy calculations along a chosen inter-molecular distance coordinate ($r$) to estimate dissociation barriers and short range PMF behavior between ion-pairs. (Figure \ref{figure:TOOL}) 
 
\subsection{A cluster model for ion-pairing}
For characterizing the local PMF along the inter-ion distance coordinate ($r$), we make several practical choices for the best compromise between efficiency and accuracy. The first choice is the level of chemical detail to be included for reasonable estimates of the PMF. For the case of ionic solutes, including explicit solvent has been shown to be essential for capturing specific ion effects\cite{muralidharan2019quasi,muralidharan2018quasi,chaudhari2020hydration} and other local effects.\cite{lytle2021lithium} This is because the physical process of dissociation of a contact ion pair (CP) to form a solvent shared/separated ion pair (SSIP/SIP) also involves the migration of solvent from outer shells into the inner solvation shell of the ions. Here, we incorporate a few layers of solvent (depending on solvent identity and density) to solvate the ions and fill the space between the ions up to a predetermined distance set to 0.65 nm. Beyond 0.65 nm the ion-pair PMF behavior is relatively flat for simple univalent ions (vide infra). Additionally, we apply a solvent barrier to prevent solvent diffusion away from the ions. We achieve this using a harmonic restraint on the number of solvents within a spherical radius, $R_{sol}$, relative to the center of distance between the ions (Figure \ref{figure:TOOL}). When $R_{sol}$ is empirically set to $2$ nm, we show that it sufficiently captures the local PMF behavior for small univalent ions relevant to battery electrolytes. 

The second choice is the level of theory required to capture the physics of ion dissociation. Ab initio molecular dynamics based on modern wave function or density functional theory approaches are accurate but computationally intensive. Classical non polarizable force-fields based on partial charges derived from quantum mechanics are more efficient and adequate for screening purposes. For instance, previous estimates of the PMF for NaCl dissociation show that force-field models capture the physics of binding accurately compared to AIMD calculations.\cite{timko2010dissociation} The method developed here is applicable for any choice of (classical or ab intio) interaction model and is expected to be computationally feasible in all cases.

The final choice is the method used for molecular dynamics sampling. Simulations based on plain vanilla molecular dynamics are useful when the barriers associated with association are small. However, enhanced sampling methods are required for evaluating PMF's in a general situation with larger barriers. Here, we choose a state-of-the-art, robust and flexible approach that incorporates an on-the-fly probability enhanced sampling (OPES) scheme.\cite{invernizzi2020rethinking} The OPES sampling scheme is able to sample all configurations necessary for characterizing the binding behavior, starting from a contact pair to a solvent separated pair in a back and forth manner until convergence of free energy. To further accelerate convergence, we restrict sampling of non-local distances ($\geq$0.65nm) by utilizing a one sided harmonic upper wall. These practical choices lead to an efficient computational model for estimating local ion-pair PMF's.

\subsection{Electrolyte screening perspective: Introducing Cluster Ion-Pair Sampling (CLIPS) Tool}

In the last few decades, modern computational approaches have played a significant role in engineering materials with desirable properties. Recent efforts such as the Materials Initiative\cite{national2011materials,jain2013commentary} seek to establish an infrastructure to accelerate advanced materials discovery by leveraging the use of computational and experimental capabilities in an integrated approach to materials engineering. In the sphere of electrolyte design, the Electrolyte Genome project\cite{qu2015electrolyte} enabled an automated high throughput computation of properties such as oxidation-reduction potential, salt solubility, and electrochemical solution stability for assessing and screening novel battery electrolytes. Further, screening electrolytes for lithium ion (Li$^+$) battery applications requires the simultaneous optimization of a variety of design parameters. Thus, a need of the hour is the development of robust, efficient models and tools to compute design parameters for supplementing such design and screening efforts.

To this end, we create a tool titled ``CLIPS'' (short for CLuster Ion-Pair Sampling) that estimates metrics related to the binding of ionic species using our simplified cluster model (Figure \ref{figure:TOOL}). This tool is made available for free on Github (\url{https://github.com/ajaymur91/CLIPS.git}). We achieve a remarkable efficiency ($\approx$ 5 minutes per PMF calculation with modest resources: 50\% of cpu resources on a single AMD Ryzen 5 3600) by combining a reduced gas-phase cluster model (containing an ion pair and a few layers of solvent) with an enhanced sampling scheme that allows efficient crossing of free energy barriers. We demonstrate how this tool can generate useful metrics related to ion correlations for screening candidates for battery electrolytes as well as poly-electrolyte solutions.

 	\begin{figure*}
 	 	\includegraphics[width=0.8\linewidth]{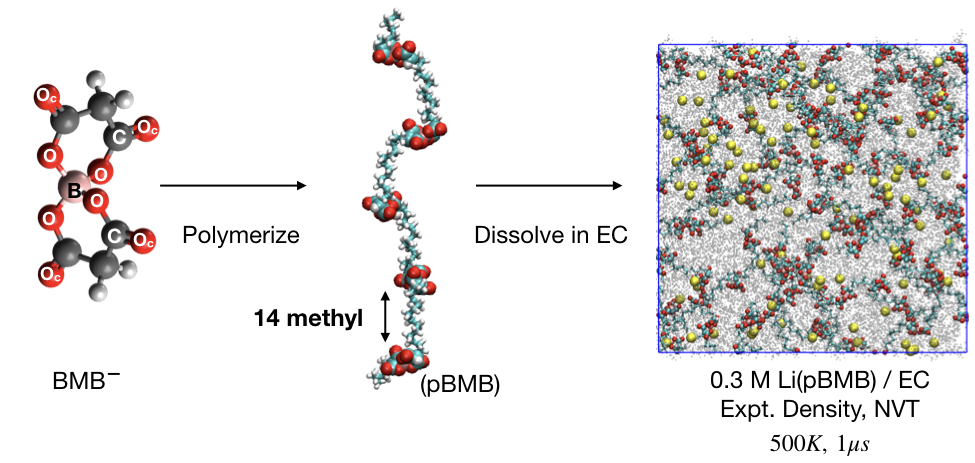}%
 	 	\caption{Simulation model for poly-electrolyte system based on 0.3 M Li(pBMB) in ethylene carbonate (EC) solvent at 500K.} 
 	 	\label{fig:pblb_model}
 	 	\end{figure*}
 	 	
\section{Computational Methods}

\subsection{Cluster model for ion-pairing}

On-the-fly Probability Enhanced Sampling Method (OPES)\cite{invernizzi2020rethinking} is a state-of-the-art method to sample systems that present with large energy barriers along slow reaction coordinates or collective variables (CV's).  We use the PLUMED\cite{bonomi2009plumed,tribello2014plumed,bonomi2019promoting} plugin with Gromacs 2019.6 for OPES simulations. We perform 5 ns of OPES at 313K in an equilibrated box of $N_s$ solvent molecules with 1 ion pair. We apply a history dependent bias, $V_n(r)$, at each step to the distance ($r$) between the cation (e.g. Li$^{+}$) and a distinguished atom of the anion (e.g. S of TFSI). 
\begin{align}
V_{n}(r) = \left( 1 - \frac{1}{\gamma}\right)\frac{1}{\beta}\log\left(\frac{P_{n}(r)}{Z_{n}} + \epsilon \right)
\label{eqn:opes}
\end{align}
where $P_n(r)$ is an estimate (at step n) of the unbiased probability distribution of $r$, and $\epsilon$ and $Z_{n}$ are the regularization and normalization terms over the explored CV-space.\cite{invernizzi2020rethinking}
We set the maximum barrier height ($\Delta E$) or the free energy barrier to overcome to $\Delta E = 100$ kJ/mol where $\epsilon = e^{-\beta \Delta E/(1-1/\gamma)}$ and $\gamma = \beta \Delta E$. This choice of $\Delta E$ is based on sampling observed in trial OPES simulations. We also limit the exploration of the cation--anion distance ($r$) by introducing a large upper wall bias according to:
\begin{align}
V_{\mathrm{wall}} =
\begin{cases}
\kappa_w (r-r_{0})^2 & \text{if $r > r_{0}$,} \\
0 & \text{if $r < r_{0}$. } \\
\end{cases}
\label{eqn:wall}
\end{align}
where $\kappa_w$ = $200000~$kJ/nm and $r_{0} = 0.65 ~$nm.

We apply a solvent barrier using harmonic restraints relative to the center of distance (c) between the ions to prevent solvent diffusion away from the ions (Figure \ref{figure:TOOL}). 
\begin{align}
V_{\mathrm{barrier}} =
\begin{cases}
\kappa_s (N-N_s)^2 & \text{if $N < N_s$} \\
0 & \text{if $N = N_s$} \\
\end{cases}
\label{eqn:barrier}
\end{align}
where $\kappa_s$ = $1000~$kJ/nm.
where N is the number of solvent molecules within a spherical radius R$_{sol}$. We choose a radius, $R_{sol} = 2~$nm, for all cases in our analysis. and $N_s=30$ or $60$ for the case of ethylene carbonate and water respectively. In general, a solvent molecule is considered to lie inside a spherical volume if a distinguished atom (i) of the solvent lies within the sphere. However, in practice we use a smooth switching function ($s_{ic}$) that goes from 0 to 1 based on the distance of the distinguished solvent atom i from $c$.  
\begin{align}
\text{N} &= \sum_{i \in solv} s_{ic}
\end{align}
\begin{align}
s_{ic} &= \frac{1 - \left(\frac{r_{ic}}{R_{sol}}\right)^n}{1 - \left(\frac{r_{ic}}{R_{sol}}\right)^m}
\end{align}
We make standard choices of $n=6$, $m=2n$ for rational switching functions. $R_{sol} = 2~$ nm.

Finally, we obtain an estimate for the unbiased distribution for the cation-anion distance, $r$, by using a re-weighting scheme\cite{tiwary2015time}:
\begin{align}
\left \langle r \right \rangle_{0} &= \frac{\left \langle r ~ e^{\beta V_T}\right \rangle_{V_T}}{\left \langle e^{\beta V_T}\right \rangle_{V_T}}.
\label{eqn:reweight1}
\end{align}

Here, $\left \langle \right \rangle_{0}$ represents an ensemble average over an unbiased trajectory, while $\left \langle \right \rangle_{V_T}$ represents a biased trajectory. $V_T$ is the sum total of the deposited bias after time $t$, including bias due to OPES, upper wall and solvent barrier (eq. \ref{eqn:opes},\ref{eqn:wall},\ref{eqn:barrier}). From the unbiased distribution, $\left \langle r\right \rangle_{0}$, we obtain the unbiased free energy as $-\log P(r)$. Further, we add a Jacobian correction\cite{chipot2007free,trzesniak2007comparison} of $2kT\ln r$ to the distance dependent free energies estimated from the unbiased probability distribution (eq. \ref{eqn:reweight1}). 

\subsection{Implementation of cluster model}
We use Gromacs 2019.6\cite{abraham_gromacs_2015,berendsen_gromacs_1995,hess_gromacs_2008,lindahl_gromacs_2019,pall_tackling_2015,pronk_gromacs_2013,van_der_spoel_gromacs_2005} with PLUMED\cite{bonomi2009plumed,tribello2014plumed,bonomi2019promoting} plugin to perform OPES simulations of NaCl (in H$_2$O) and some common Li$^+$ battery electrolytes. Lithium bis(trifluoromethane) sulfonimide (LiTFSI) and Lithium trifluoromethanesulfonate (LiOTf) in ethylene carbonate (EC) are ideal test cases because they display distinct ion pairing behavior despite the anions being structurally similar.\cite{lytle2021lithium} Force field parameters for Li$^+$, OTf$^-$, TFSI$^-$ and ethylene carbonate are taken from our previous work.\cite{lytle2021lithium} OPLSAA parameters \cite{canongia_lopes_molecular_2004} are used for Na$^+$ and Cl$^-$ with SPC water model.\cite{berendsen1981interaction} All bonds are constrained using the LINCS algorithm. Electrostatic interactions are calculated using the particle mesh ewald technique with a 1.2 nm cutoff. A cut-off Lennard-Jones interaction is implemented for the non-electrostatic interactions with a 1.2 nm cutoff. Simulation systems undergo energy minimization using the steepest descent minimization algorithm. After energy minimization, 5 ns of OPES is performed with the stochastic dynamics\cite{goga2012efficient} integrator at 313K in a simulation box with a few layers of solvent and 1 ion pair. Our implementation is made available on Zenodo\cite{muralidharan_ajay_2021_5559115} and Github.
 	 	
\subsection{poly-electrolyte simulations}

We perform a $1 \mu s$ NVT simulation (Figure \ref{fig:pblb_model}) of 0.3M of Li$^+$ and poly(bis(nonenylmalonato)-borate) salt in bulk ethylene carbonate (EC) solvent (0.3 M Li(pBMB)/EC) at 500K. The poly-anion model (pBMB) is built by combining a force-field model\cite{wang2014atomistic} for BMB$^-$ with a model for alkyl (methyl) spacer groups.\cite{siu2012optimization}.  The simulation cell contains 100 Li$^+$ and 20 poly-ions (5-mers) packed randomly in a cubic box of length $8.2~nm^3$ with 4500 ethylene carbonate (EC) solvent molecules. After energy minimization using a steepest descent algorithm, we carry out an initial $1 \mu s$ of NVT equilibration with the Nose-Hoover thermostat\cite{nose1984molecular,hoover1985canonical} at a density of $1320~kg/m^3$ followed by a $1 \mu s$ molecular dynamics trajectory that is used for further analysis.
 	 	
We perform cluster simulations to characterize the PMF between Li$^+$ and  bis(nonenylmalonato)-borate (BMB$^-$) anion in EC solvent.  We label the partial charge distribution from the model of Wang \textit{et al.} \cite{wang2014atomistic} as Q1 (inset, Figure \ref{fig:poly}). We also create an alternate hypothetical de-localized charge model for BMB$^-$ anion labeled as Q2. Note that we obtain Q2 model (from Q1) by manual redistribution of atomic partial charges (inset Figure \ref{fig:poly}). 

 \begin{figure}
 	 	\includegraphics[width=\linewidth]{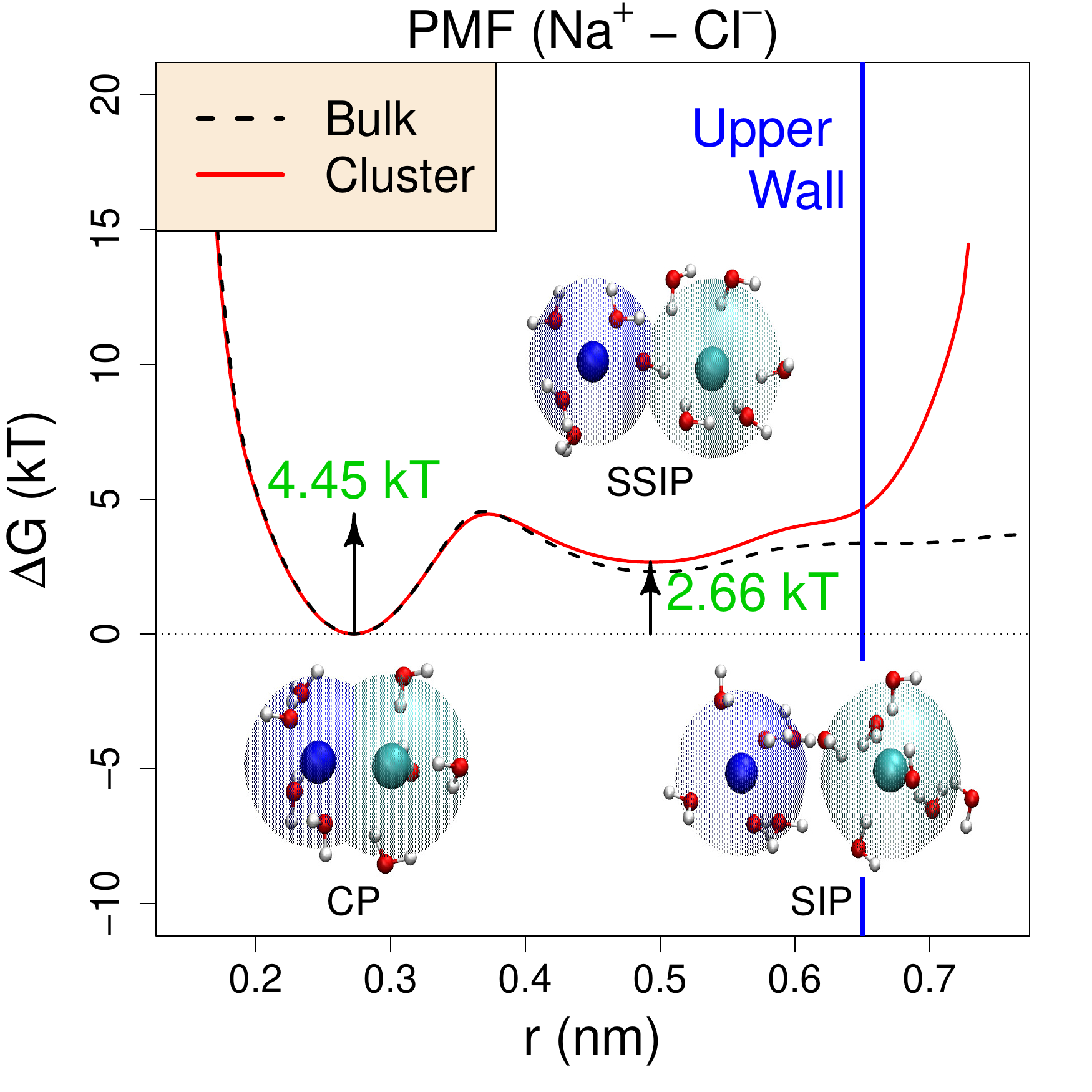}%
 	\caption{Potential of mean force (PMF) for NaCl in water from bulk solvent simulations (black dashed) and the cluster model (red) along the Na$^+$--Cl$^-$ distance. The arrows indicate the dissociation barrier and the free energy difference between the contact pair (CP) and solvent shared ion pair state (SSIP). The blue vertical line locates the harmonic upper wall bias that prevents exploration of larger distances ($\geq 0.65 ~nm$) between the ions. .}%
 	\label{figure:PMF1}
 \end{figure}

\begin{figure*}[th!]
  	\includegraphics[width=0.45\linewidth]{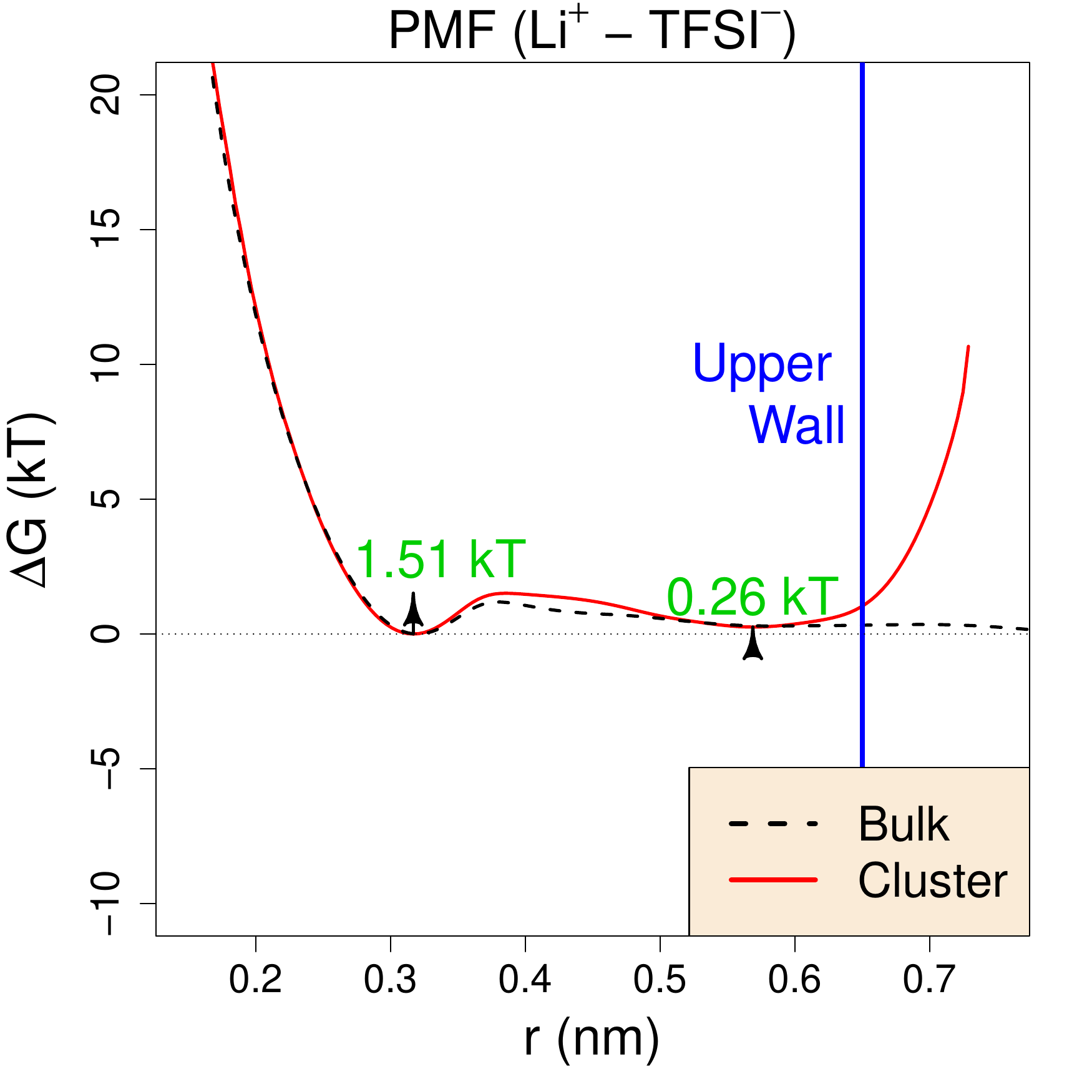}%
 	 	 	 	\hfill
 	\includegraphics[width=0.45\linewidth]{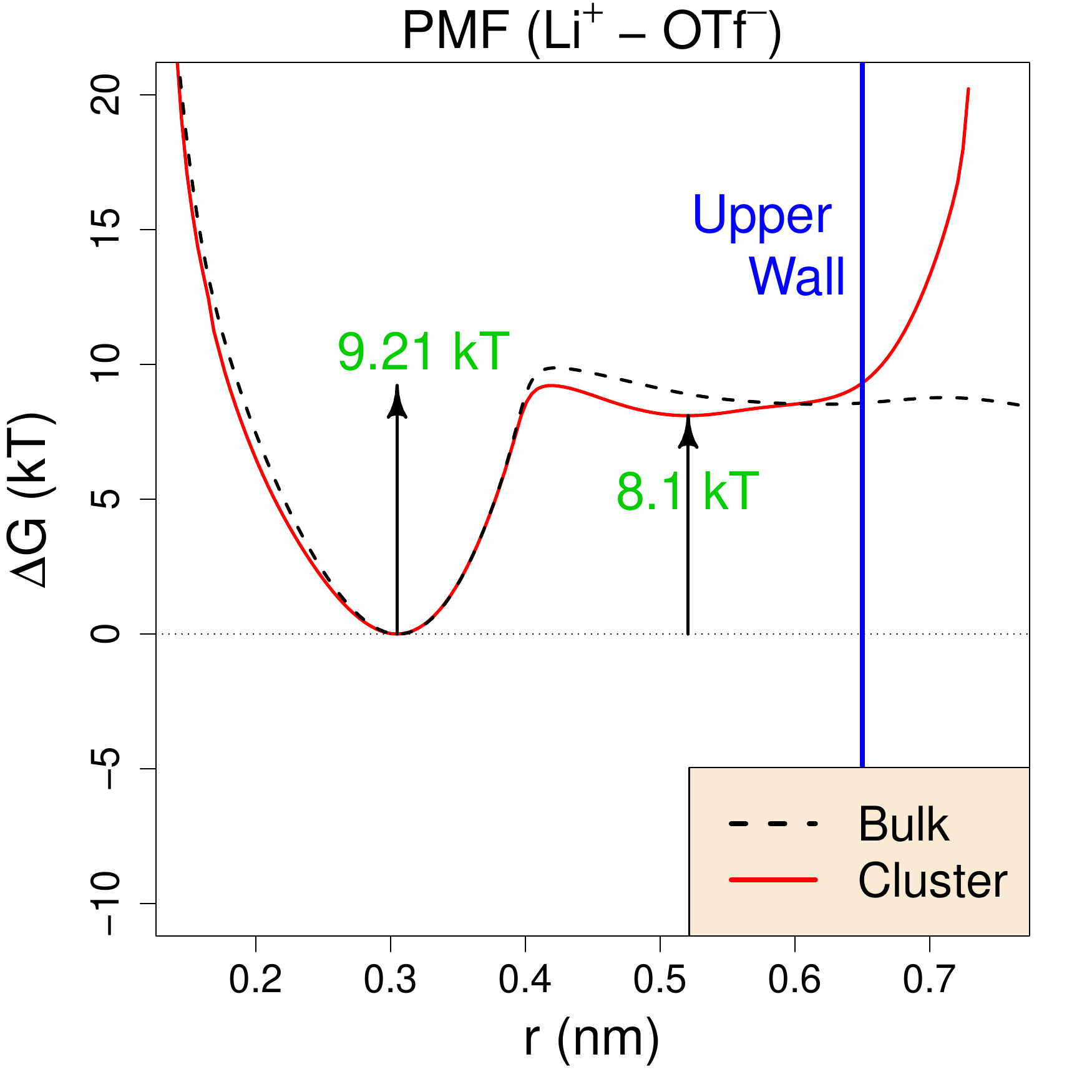}%
 
 	\caption{Potential of mean force (PMF) for LiTFSI and LiOTf from the cluster model (red) compared to results from bulk simulations \cite{lytle2021lithium} (dashed black lines).}
 	\label{figure:PMF2}
 \end{figure*}
\section{Results and Discussion}
\label{sec:results}

Our cluster model with enhanced sampling (Figure \ref{figure:TOOL}) approximates bulk solvent ion-pairing potential of mean force (PMF).
The cluster simulations begin with the ions in the contact pair (CP) state.  Dissociation barriers are crossed as OPES biases (eq. \ref{eqn:opes}) are added, and the solvent shared ion pair (SSIP) and solvent separated ion pair (SIP) states are explored revealing the local free energy landscape. By fast sampling back and forth, we obtain very good estimates of the dissociation and association barriers and free energy differences between the CP and SIP states.  An animation of our model in action is provided on figshare (\url{https://doi.org/10.6084/m9.figshare.16755277.v3}).   The upper wall (blue vertical lines in the figures) harmonic bias prevents the exploration of distances beyond 0.65 nm and is sufficient to capture the interesting details of the local free energy landscape for small ions.  

 When compared to bulk solvent simulations, the cluster model shows good agreement for the PMF for NaCl salt in water (Figure \ref{figure:PMF1}) and LiTFSI and Li-OTf in EC (Figure \ref{figure:PMF2}).  In all cases, the agreement is excellent for distances less than the upper wall at 0.65nm.
Beyond that distance the bulk solvent free energy is relatively flat compared to smaller distances and hence not targeted by our cluster model.  For NaCl, the barrier between the CIP and SSIP is $\sim 4.55 kT$ which is accurately reproduced by the cluster model.  For Li$^+$-- OTf$^-$ that barrier is much higher $\sim 9.21 kT$, however, reproduced faithfully by the cluster model as well. The model is therefore able to capture the essential physics of ion-pairing for distinct cases of free energy barriers at a very modest computational expense.

\begin{figure*}[th!]
  	\includegraphics[width=0.45\linewidth]{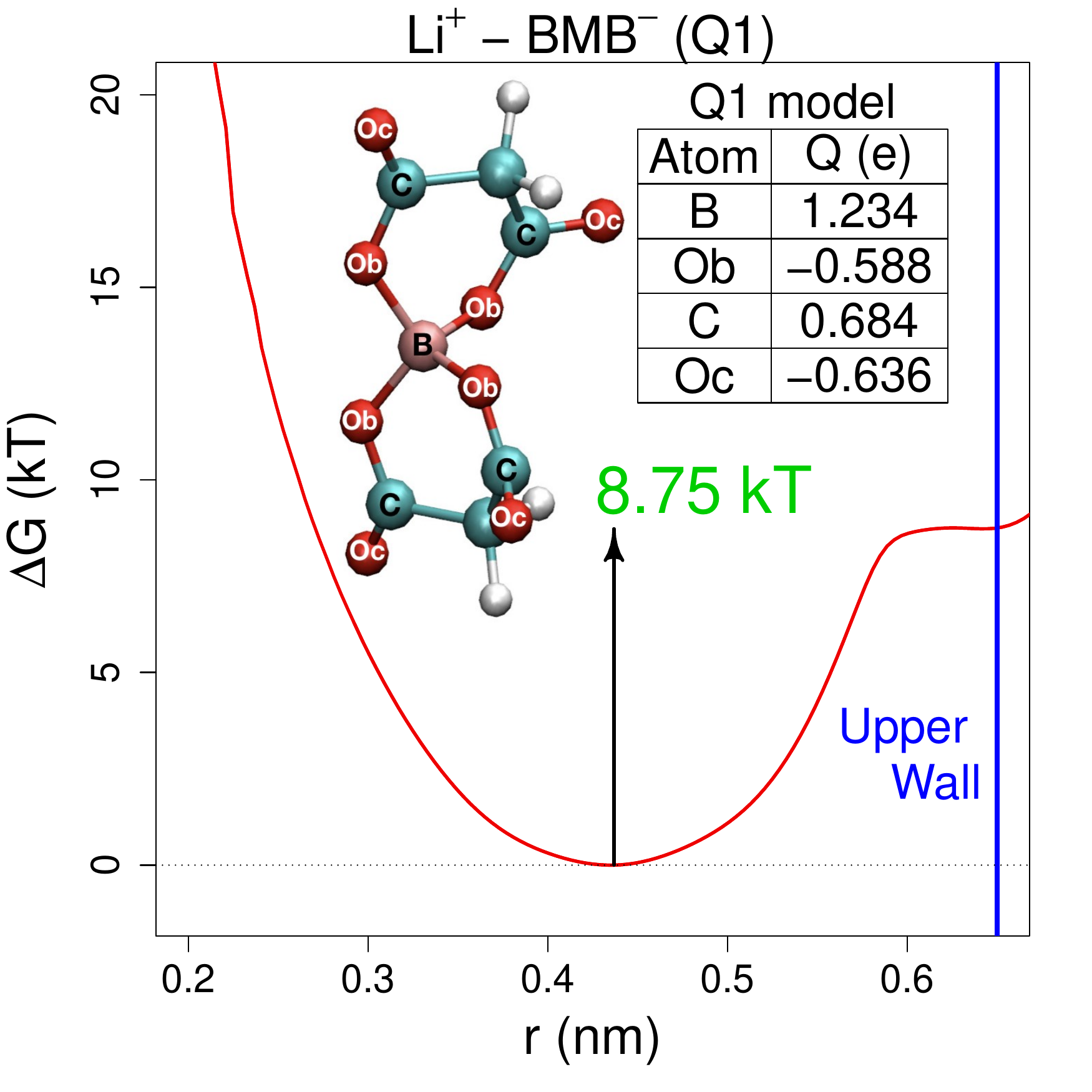}%
 	\hfill
 	\includegraphics[width=0.45\linewidth]{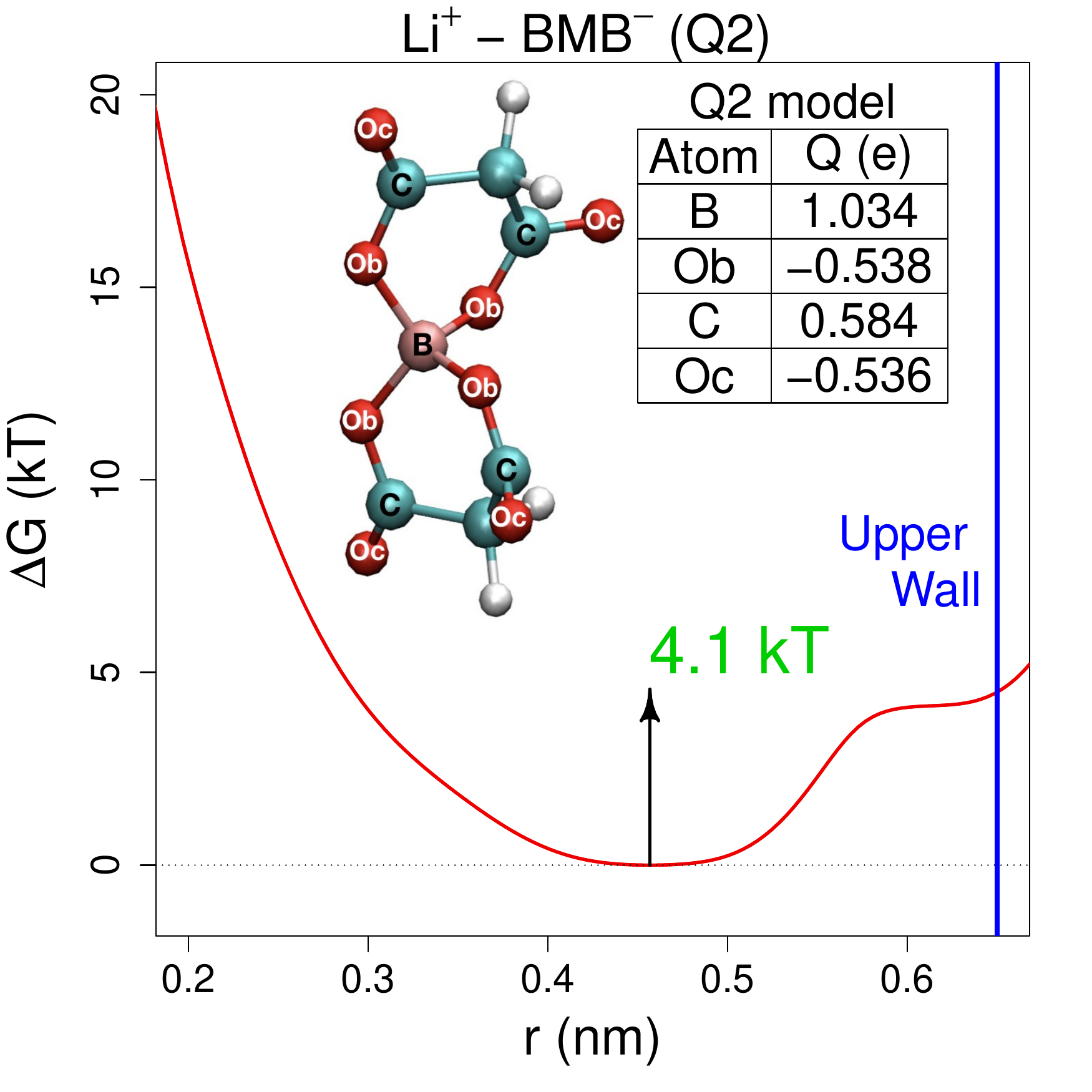}
	\caption{PMF between Li$^+$ and BMB$^-$ anion from cluster simulations for Q1 (original) and Q2 (modified/ hypothetical) charge models.}\label{fig:polypmf}
\end{figure*}

\begin{figure*}
	\hspace{0.5cm}
 	 	\includegraphics[width=0.35\textwidth,height=0.35\textwidth]{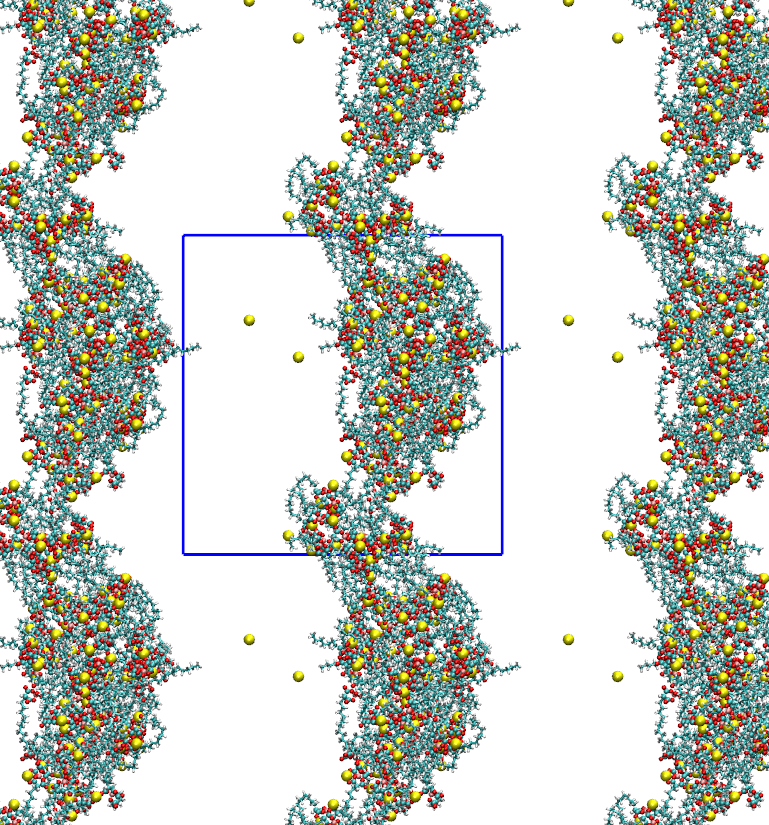}%
 	 	\hspace{3cm}
 	\includegraphics[width=0.35\textwidth,height=0.35\textwidth]{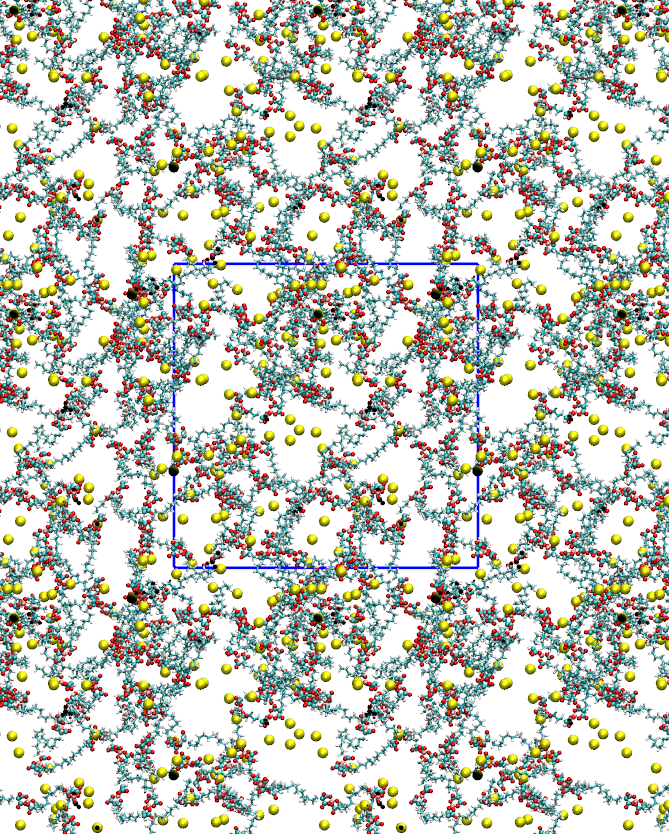}%

  	\caption{Comparison of micro-structures of 0.3 M p(LiBMB) in bulk ethylene carbonate (EC) solvent at 500K for Q1 (left) and Q2 (right) based charge model for the poly-anions. The solvent is not shown here for clarity.}%
 	\label{fig:poly}
 \end{figure*}

In light of the successful treatment of PMF's for simple salts with our cluster model, we shift our focus to the case of poly-electrolytes. Recent experiments\cite{dewing_electrochemically_2020} have sought to develop a polymer solution electrolyte based on poly(bis(nonenylmalonato)-borate) [Li(pBMB)] that is electrochemically stable over a wide potential window in organic carbonate solvents with high Li$^+$ transference for batteries. However, in practice the poly-electrolyte [Li(pBMB)] appears to make a gel in EC solvent the mechanism for which is not clear.  Figure~\ref{fig:polypmf} depicts the PMF between Li$^+$ and BMB$^-$, and shows that there is a deep attractive well which can result in strong ion pairing. This calculation was based on the force field obtained from Wang \textit{et al}.\cite{wang2014atomistic} which we label as Q1.  From molecular dynamics simulations of the solution, we find that in this case there is an aggregation of ions that persists over the duration of the simulation.  Figure~\ref{fig:poly} (left panel) shows the final configuration from  molecular dynamics simulations at experimentally relevant\cite{dewing_electrochemically_2020} concentration of 0.3M Li(pBMB) in EC solvent. The degree of polymerization of the poly-anion is 5. The simulations reveal that the system based on the Q1 (strong interaction) model for BMB$^-$ formed non-homogeneous micro porous structures with self diffusivities of $0.24~$x$~10^{-6}$ cm$^2$/s for Li$^+$ and $0.05~$x$~10^{-6}$ cm$^2$/s for monomer beads of BMB$^-$.

The undesirable gelation behavior of the poly-electrolyte can be reduced dramatically with subtle changes in the charge distribution of the anion.  To test the effect of charge delocalization we adjust the partial charges in the Q1 force-field by hand.  In the resulting Q2 model, the partial charges on O$_\mathrm{b}$ and O$_\mathrm{c}$ oxygen atoms are reduced by 8\% and 15\% (inset) respectively compared to the Q1 model. The charges on boron and carbonyl carbon are adjusted to keep the net charge of Q2 model to be -1. With these slight adjustments the PMF well depth decreases by more than a factor of two (see Figure~\ref{fig:polypmf} ) in cluster simulations.   This difference has a dramatic influence on the micro-structure, and the weak association (for Q2 model) between Li$^+$ and BMB$^-$ in the solution phase leading to homogeneous micro-structure with better Li$^+$ transport (see Figure \ref{fig:poly}).  The Q2 based poly-electrolyte model produces improved self diffusivities of $5.22~$x$~10^{-6}$ cm$^2$/s for Li$^+$ and $0.57~$x$~10^{-6}$ cm$^2$/s for monomer beads of BMB$^-$. 

\section{Summary and conclusions}

Ion-pairing can determine the viscosity and ionic transport coefficients in electrolyte solutions.  An intrinsic propensity for ion pairing is the PMF between ions in dilute solution. We introduce a robust and efficient cluster model to estimate this PMF. To achieve high efficiency, the model combines enhanced sampling OPES with a cluster of explicit solvent molecules. The predictions of the model are in excellent agreement with bulk solvent simulations. 

We use the model to predict the PMF between  Li$^+$ and the complex anion BMB$^-$ in EC.  The calculations show a deep attractive well that should lead to ion-pairing in dilute solution.  From molecular dynamics simulations of the bulk system we find that there is an aggregation of ion pairs and a gel-like micro-structure.  We find that subtle changes in the charge distribution of the anion (without changing the overall charge of course) results in a large difference in the PMF, and the micro-structure of the bulk liquid, which becomes homogeneous.  The ion self-diffusion coefficients increase by an order of magnitude.   The calculations therefore provide a design principle for anion selection.  Judicious modification of the anion chemistry to make the charge more diffuse might result in better structural and transport properties for the poly-electrolyte. 

In concentrated solutions, especially with divalent ions, aggregation of ions might play an important role.  In this case, in addition to the pair PMF the free energy of multiple ions could be of importance.  The cluster model can be extended to such systems with larger clusters of solvent molecules, and an appropriate choice of collective variables, i.e., order parameters.  Investigating many body effects is a possible future direction of this work.

An interesting aspect to consider is the the number of solvent molecules required to reproduce the PMF obtained from bulk solvent simulations\cite{lytle2021lithium}.  In all our calculations, we empirically set the number of solvent molecules to completely cover Li$^+$ and the space between the ions at all ionic separations up to 0.65 nm.  It might he possible to optimize this because in practice we are not interested in the long-range behavior of the PMF, but rather the short-ranged behavior where ion-pairing occurs.  Another possibility is to use very few solvent molecules, and a reaction field.  

The method is easily extensible to more realistic interactions between the ions and solvent molecules.  Our CLIPS tool performs calculations at the classical force-field level because highly automated implementations of accurate force-fields are available (OPLSAA Ligpargen,\cite{canongia_lopes_molecular_2004,dodda2017ligpargen} CHARMM\cite{vanommeslaeghe2010charmm}) with parameters derived from quantum mechanics. However, our model is general and independent of the level of theory used for performing the simulations. For instance, if force-fields are not available, ab initio methods such as density functional theory (DFT) or semi-empirical methods such as Density Functional Based Tight Binding (DFTB) can be used instead. The significance of our cluster model becomes particularly evident for ab initio based computations because bulk solvent trajectories are prohibitively expensive.

\section*{Supplementary Info}
\begin{itemize}
    \item Cluster Ion-Pair Sampling (CLIPS) tool:\\
\url{https://github.com/ajaymur91/CLIPS.git}
    \item Animation of tool:
\url{https://doi.org/10.6084/m9.figshare.16755277.v3}
\end{itemize}
\begin{acknowledgements}
This research was supported in part by UW-Madison Department of Chemistry PHOENIX \& STARLING research cluster through National Science Foundation Grant CHE-0840494. This study was also partially supported by US Department of Energy, Basic Energy Sciences Contract DE-SC0017877. A.M. is a Hirschfelder fellow. 
\end{acknowledgements}
\clearpage
\nocite{*}
\bibliography{clips_bib.bib}

\appendix
\section{Table of contents graphic}
\includegraphics[width=0.5\textwidth]{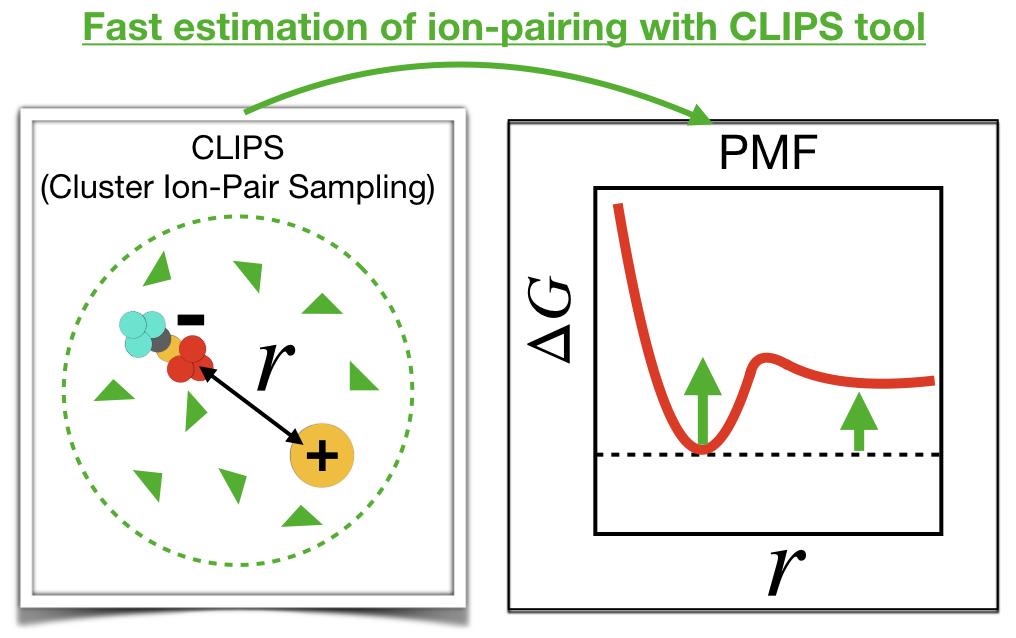}
\end{document}